\documentstyle[12pt,aps,floats,prc,epsfig,tighten]{revtex}
\def\Journal#1#2#3#4{{\em #1} {\bf #2}, #3 (#4) }
\def\NPA{{ Nucl. Phys.} {\bf A}}
\def\PRL{Phys. Rev. Lett.}
\def\PREV{ Phys. Rev.}
\def\PREP{ Phys. Rep.}
\def\PRC{{Phys. Rev.} C}
\def\PL {Phys. Lett.}
\begin{document}
\title{ Nuclear Self-energy and Realistic Interactions}
\author{T. Frick, Kh. Gad,  H. M\"uther}
\address{Institut f\"ur
Theoretische Physik, \\ Universit\"at T\"ubingen, D-72076 T\"ubingen, Germany}
\author{ P. Czerski}
\address{Institute Fiz.~Jadrowej, Pl-31-342 Krakow, Poland}
\maketitle
 
\begin{abstract} 
The structure of nucleon self-energy in nuclear matter is evaluated for various
realistic models of the nucleon-nucleon (NN) interaction. Starting from the
Brueckner-Hartree-Fock approximation without the usual angle-average
approximation, the effects of hole-hole contributions and a self-consistent
treatment within the framework of the Green function approach are investigated.
Special attention is paid to the predictions for the spectral function
originating from various models of the NN interaction which all yield an
accurate fit for the NN phase shifts. 
\end{abstract}                    
\section{Introduction}
One of the central issues of quantum many-body theory and theoretical nuclear
physics is the attempt to derive the bulk properties of nuclear systems from   
realistic models of the nucleon-nucleon (NN) interaction. Various approximation
schemes have been developed to describe the correlations which are induced into
the many-nucleon wave function by the strong short-range and tensor components
of a such a realistic NN force. For a recent review on such methods 
see e.g.~\cite{bookbal,mupo}. One of the most popular approximation schemes,
which is frequently used in nuclear physics, is the hole-line expansion, in
particular the approximation of lowest order, the Brueckner-Hartree-Fock (BHF)
approximation. 

In this approach the NN correlation are taken into account in solving 
the Bethe-Goldstone equation, which leads to the so-called G-matrix. The
G-matrix accounts for multiple scattering processes and therefore corresponds 
to the NN scattering matrix $T$. In contrast to the Lippmann-Schwinger equation
leading to $T$ the Bethe-Goldstone equation accounts for effects of the
nuclear medium: The propagator for the intermediate two-particle states is
restricted to particle states, i.e. to single-particle states with energies
above the Fermi energy $\varepsilon_F$, and is defined in terms of the
single-particle energies for the nucleons in the medium. This choice of the
single-particle energies in the propagator of the Bethe-Goldstone equation is
motivated by the Bethe-Brandow-Petschek theorem\cite{BBP}.  

Strictly speaking, however, the Bethe-Brandow-Petschek theorem only defines the
energy variable to be used in the calculation of self-energy or single-particle
potential for the hole states. The choice for the propagator of the particle
states is not defined on this level of the hole-line expansion and therefore has
been discussed in a controversial way. The conventional choice has been to ignore
self-energy contributions for the particle states completely and approximate the
energies by the kinetic energy only. This conventional choice is supported by
the $S_2$ approximation of the coupled cluster or exponential S
method\cite{kuem}, which essentially leads to the same approach,  This
conventional choice for the single-particle spectrum, however, is not very
appealing as it leads to a gap at the Fermi surface: the propagator for 
single-particle states with momenta below the Fermi momentum $k_F$ is described
in terms of a bound single-particle energy while the corresponding spectrum for
the particle states starts at the kinetic energy for the momentum $k_F$.

Mahaux and collaborators\cite{jeuk} argued that it would be more natural to
choose the propagator according to the Green function method, i.e.~define the
single-particle propagator with a single-particle energy which includes the real
part of the self-energy as a single-particle potential for particle and hole 
states. This leads to a spectrum which is continuous at the Fermi momentum, which
provided the name ``continuous choice'' for this approach. This continuous choice
leads to an enhancement of correlation effects in the medium and tends to
predict larger binding energies for nuclear matter than the conventional choice.

The optimal choice for the single-particle spectrum seems to be in between the
conventional and the continuous choice spectrum. This has been demonstrated by
the recent investigations of the Catania group\cite{baldo}. They calculated the
effects of contributions in the hole line expansion originating from three-hole
line terms\cite{day} using both the conventional and the continuous choice
spectrum. The inclusion of the three-hole line terms leads to a calculated
binding energy of nuclear matter which is in between the results of BHF
calculation with the conventional and the continuous choice and rather
insensitive to the choice made. The attractive contribution obtained from the
three hole line terms using the conventional choice is  larger than the
repulsive contribution evaluated with the continuous choice. This demonstrates
that the continuous choice yields a slightly better convergence than the
conventional one. The results also indicate that the two-nucleon correlations
tend to be underestimated with the conventional choice and slightly
overestimated with the continuous choice for the single-particle spectrum. It is
one of the aims of this manuscript to explore the role of the single-particle
propagator in the evaluation of NN correlation in more detail.

In solving the Bethe-Goldstone equations and its extensions to the three hole
line terms, it has been common practise to use an approximation for the Pauli
operator which is independent on the angle between the center of mass momentum
of the interacting pair of nucleons and its relative momentum. Using the
angle-average approximation and assuming that the single-particle spectrum can
be parametrised as function of the momentum in a quadratic form, one can solve
the Bethe-Goldstone equation easily using a partial wave expansion. During the
last few years methods have been developed to solve the Bethe-Goldstone equation
without this angle-average approximation\cite{schiller,suzuk}. It turns out
that the results are not very sensitive to the treatment of the Pauli operator
but they change in a non-negligible way if a single-particle spectrum is used,
which differs from a simple quadratic parametrisation. This has recently been
observed by Baldo and Fiasconaro\cite{fiasco} using an angle-average
approach. We are going to explore this feature in more detail without using the
angle-average approximation.

The continuous choice for the single-particle spectrum has been motivated from
the method of self-consistent Green functions in the many-body theory.
Within this scheme, however, the evaluation of the
self-energy is done treating particle-particle and hole-hole ladders at the
same level of the approximation scheme replacing the Bethe-Goldstone equation
by the corresponding Galitzkii-Feynman equation. If one tries to perform such
calculations for realistic NN forces, one encounters the problem of the
so-called pairing instabilities\cite{vonder,alm,bozek}. 

These pairing effects can be taken into account by means of the BCS
approach\cite{baldbcs,almbcs,elgar}. At the empirical saturation density of
symmetric nuclear matter the solution of the gap equation in the $^3S_1-^3D_1$
partial wave leads to an energy gap of around 10 MeV. Another approach is 
to consider an evaluation of the generalised ladder diagrams with ``dressed'' 
single-particle propagators. This means that the single-particle Green functions
are not approximated by a mean-field approach but consider single-particle
strength distributed over all energies. Various attempts have been made in this
direction, considering a parametrisation of the single-particle Green function
in terms of various poles\cite{dimitr}, employing simplified (separable)
interaction models\cite{bozek} or considering the case of finite temperature.

The same instabilities also occur in studies of finite nuclei\cite{heinz},
leading to divergent contributions to the binding energy from the generalised
ring diagrams. These contributions remain finite if the single-particle
propagators are dressed in a self-consistent way.

Here we want to explore some features of the single-particle self-energy and the
corresponding Green function for infinite symmetric nuclear matter at temperature
$T=0$. As a starting point we consider the BHF approximation employing a
self-consistent continuous choice spectrum without angle-averaging. The
contribution of hole-hole ladder terms are then added in a perturbative
approach. We then analyse the energy distribution of the spectral
single-particle strength and define appropriate mean values. These mean values
are used to calculate the self-energy in a self-consistent way. 

These studies are performed for various NN interactions and we will pay special
attention to the differences in the properties of the nucleon self-energy which
can be related to the interaction model. In particular, we will consider local
and non-local models of the NN interaction, which have  been defined by the
groups in Moscow (Idaho)\cite{cdb}, Argonne\cite{argv18}, and
Nijmegen\cite{nijm1}. All these interaction models yield a very accurate fit to
a selected data-base of NN scattering data. For further comparison we will also
consider two older version of the Bonn potential, defined in \cite{rupr}.

After this introduction we will discuss some features of the self-consistent 
BHF approach for the self-energy in section 2. The effect of the hole-hole terms
in the self-energy and a self-consistent treatment of the single-particle Green
function will be presented in section 3. The main conclusions are summarised in
the final section.

\section{Single Particle Spectrum in the BHF Approximation}

The self-energy or single-particle energy of a particle in the 
Brueckner-Hartree-Fock (BHF) approximation corresponds to the Hartree-Fock
expression using the $G$ matrix for the effective interaction. This means that 
the self-energy of a nucleon in nuclear matter with momentum $\vec k$ is given
by (note that spin and isospin quantum numbers are suppressed)
\begin{equation}
\Sigma_{BHF} (\vec k,\omega ) = \int d^3q <\vec k\vec q | G(\Omega ) |\vec k
\vec q > n_0(\vec q) \,,\label{eq:selfbhf}
\end{equation}
with the occupation probability of a free Fermi Gas with a Fermi momentum $k_F$
\begin{equation}
n_0 (\vec q) = \left\{ \begin{array}{ll} 1 & \mbox{for}\; |\vec q| \leq k_F \\
0 & \mbox{for}\; |\vec q| > k_F \end{array}\right.\label{eq:occ0}
\end{equation}
The matrix elements in (\ref{eq:selfbhf}) denote antisymmetrized matrix elements
of the Brueckner $G$ matrix which are determined by solving the Bethe-Goldstone
equation for a given realistic nucleon nucleon (NN) interaction $V$
\begin{eqnarray}
<\vec k\vec q | G(\Omega ) |\vec k\vec q > & = <\vec k\vec q | V |\vec k\vec q > 
+ & \int d^3p_1 d^3p_2 <\vec k\vec q | V |\vec p_1 \vec p_2 >\nonumber \\ &&
\times
\frac{Q(p_1p_2)}{\Omega - (\tilde\varepsilon_{p1} + \tilde\varepsilon_{p2}) + 
i\eta}
<\vec p_1 \vec p_2| G(\Omega ) |\vec k\vec q > \,.\label{eq:betheg}
\end{eqnarray}
In this representation of the Bethe-Goldstone equation $Q$ stands for the Pauli
operator and is defined by
\begin{equation}
Q(p_1p_2) = (1 - n_0(p_1))(1-n_0(p_2)) \,, \label{eq:pauli}
\end{equation}
which means that the integral over intermediate two-particle states in
(\ref{eq:betheg}) is restricted to states with momenta larger than the Fermi
momentum. According to the theorem of Bethe, Brandow and Petschek (BBP
theorem)\cite{BBP} one defines the single-particle energy for hole
states, i.e.~states with momenta $k<k_F$, using the on-energy shell value for
the self-energy, which is given as
\begin{equation}
\varepsilon_k = \frac{k^2}{2m} +  \Sigma_{BHF} (\vec k,\omega = \varepsilon_k)
\,,\label{eq:bhf1}
\end{equation} 
with a value for the starting energy parameter $\Omega$ in the Bethe-Goldstone
equation (\ref{eq:betheg}) of
\begin{equation} 
\Omega = \omega + \varepsilon_q = \varepsilon_k + \varepsilon_q\,.
\label{eq:bhf2}
\end{equation}
Since this BBP theorem does not apply for the definition of the single-particle
energies for states with momenta above the Fermi momentum, the optimal choice
for this particle-state spectrum has been a subject of
controversial discussions for many years (see the introduction above). Note that
the binding energy per nucleon is evaluated as
\begin{equation}
\frac{E}{A} = \frac{\int d^3k\,n_0(k)\frac{1}{2}\left(\frac{k^2}{2m} + 
\varepsilon_k \right)}{\int d^3k\,n_0(k)}\,,\label{eq:ebhf}
\end{equation}
which implies that the definition of the  particle-state spectrum affects the
calculated binding energy only via the energy parameter
$\tilde\varepsilon_{p1}$ and $\tilde\varepsilon_{p2}$ in the Bethe-Goldstone
equation (\ref{eq:betheg}). Two choices have been discussed: the so-called
``conventional'' choice, in which the energies for the particle states have
been replaced by the kinetic energy
\begin{equation}
\tilde\varepsilon_{p} = \frac{p^2}{2m} \,,\label{eq:conv}
\end{equation}
and the ``continuous'' choice, for which the on-shell definition of the
hole-state energies (\ref{eq:bhf1}) has been extended to the particle states.
Note that the BHF self-energy for the continuous choice develops an imaginary
part for energies $\omega > \varepsilon_F$ with $\varepsilon_F$ denoting the
Fermi energy, the single-particle energy for $k=k_F$. The ``continuous''
choice, however, employs the real part of this self-energy, only.

The on-shell definition of the BHF self-energy in eqs.(\ref{eq:bhf1}) and  
(\ref{eq:bhf2}) implies a self-consistent solution of the Bethe-Goldstone
equation (\ref{eq:betheg}) and the evaluation of the single-particle energies.
In order to obtain such a self-consistent solution one often assumes a quadratic
dependence of the single-particle energy on the momentum of the nucleon in the
form
\begin{equation}
\varepsilon_k \approx \frac{k^2}{2m*} + C \,. \label{eq:param1}
\end{equation}
Starting with an appropriate choice for the parameters for the effective mass
$m^*$ and the constant $C$, one can solve the Bethe-Goldstone equation and
evaluate the single-particle energy, using (\ref{eq:bhf1}), for two
representative momenta $k_1$ and $k_2$. The parameters $m^*$ and $C$ are
readjusted in such a way that the parametrisation (\ref{eq:param1})
reproduces these two energies. This procedure is iterated until a
self-consistent solution is obtained.

The parametrisation of (\ref{eq:param1}), however, is useful not only to
simplify the self-consistent solution of the BHF equations, it also leads to a
simplification of the numerical solution of the Bethe-Goldstone equation.
Assuming such an effective mass spectrum, one can easily rewrite the energies
occurring in the denominator of the two-particle propagator in terms of the center
of mass, $P_{CM}$ and relative momentum $p_r$ of the interacting pair of 
nucleons
$$
\varepsilon_{p1} + \varepsilon_{p2} = \frac{P_{CM}^2}{4m^*}+\frac{p_r^2}{m^*} +
2C\,,
$$
which does not depend on the angle between $\vec P_{CM}$ and $\vec p_r$. If
furthermore one approximates the Pauli operator of (\ref{eq:pauli}) by taking an
appropriate average over this angle, the Bethe-Goldstone equation can be
rewritten into a one-dimensional integral equation using the partial wave
representation of the two-body states\cite{haftel}.

This approximation scheme has been common practise and only recently attempts
have been made to avoid the angle-average approximation for the Pauli operator
\cite{schiller,suzuk}. These investigations show that the use of the
angle-average approximation in the Pauli operator leads to an underestimation of
the calculated energies of around 0.5 MeV per nucleon if one uses an effective
mass parametrisation for the single-particle energies in the Bethe-Goldstone
equation. 

\begin{figure}
\begin{center}
\epsfig{figure=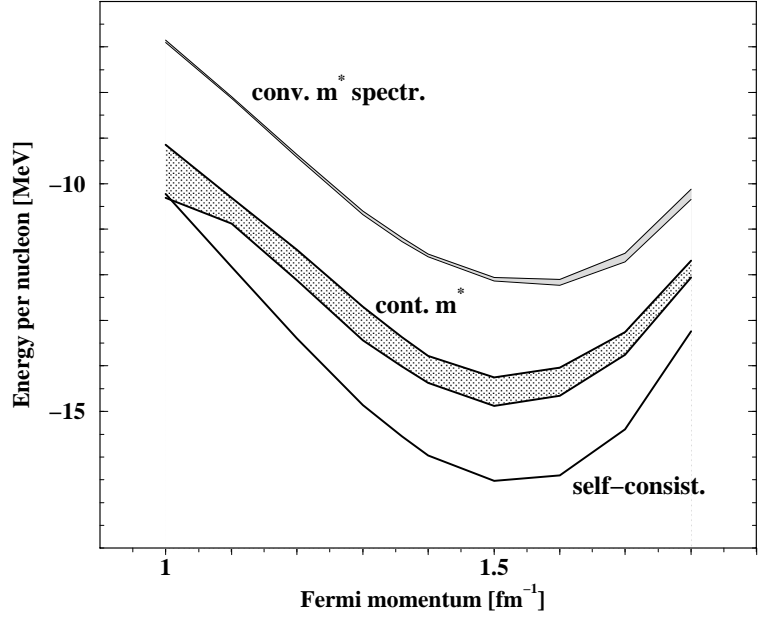,width=10cm}
\end{center}
\caption{Binding energy of nuclear matter as a function of the Fermi momentum.
Results are given for the BHF approximation using the Bonn C potential as
defined in \protect\cite{rupr}. Various approximation schemes have been used for
the two-particle propagator in the Bethe-Goldstone equation, as discussed in the
text. \label{fig1}}
\end{figure}

However, the determination of this parametrisation is not very well defined.
Rather different values for $m^*$ and $C$ may be obtained if different momenta
are chosen to define these values. This is demonstrated in Fig.~\ref{fig1},
which displays the energy of symmetric nuclear matter at various densities,
represented by the corresponding Fermi momentum. All calculations have been
performed within the framework of the BHF approach but using various
approximation schemes for the propagator in the Bethe-Goldstone equation.     
Calculations using the conventional choice for the particle-state spectrum
(\ref{eq:conv}) and the angle-average Pauli operator yield results, which are
not very sensitive to the details of the effective mass parametrisation for the
hole state energies. All resulting binding energies, which were obtained by
adjusting the parameters $m^*$ and $C$ at different momenta, are very similar
forming the thin gray area, which is labelled ``conv. $m^*$ spectr.'' in
Fig.~\ref{fig1}. 

The calculated binding energies are larger, if the continuous choice is employed,
and the results are more sensitive to the details of the procedure to determine
the parameters $m^*$ and $C$. This is visualised by the gray area, labelled
``cont. $m^*$'' in Fig.~\ref{fig1}. Both of these features can be related to the
fact that the continuous choice spectrum tends to lead to energy denominators in
the Bethe-Goldstone equation (\ref{eq:betheg}) with smaller absolute values than
the conventional choice, which exhibits a gap in the single-particle spectrum at
$k=k_F$.

The calculated binding energies are even larger, if the exact Pauli operator is
employed and the parametrisation of the single-particle spectrum is avoided
(solid line in Fig.~\ref{fig1}). The results displayed in Fig.~\ref{fig1} have
been obtained employing a specific version of the Bonn potential (Bonn C as
defined in \cite{rupr}). Very similar results have also been obtained using other
realistic models for the NN interaction. A
part of this additional binding energy can be related to the use of the exact
Pauli operator (see Ref.\cite{schiller} and discussion above). Another part,
however, must be related to the fact that the calculated single-particle
spectrum deviates in a significant way from the parametrisation of
(\ref{eq:param1}). Such a gain of binding energy has also recently been observed
by Baldo and Fiasconaro\cite{fiasco} using an angle-average approach.

\begin{figure}
\begin{center}
\epsfig{figure=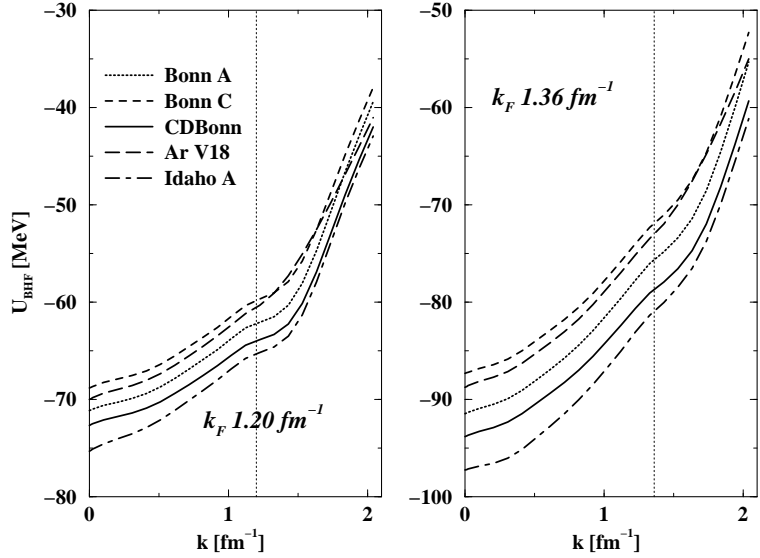,width=10cm}
\end{center}
\caption{The single-particle potential $U_{BHF}$ (see
eq.(\protect\ref{eq:ubhf}))
as a function of the momentum. Results are displayed at two different densities
using various models for the NN interaction.\label{fig2}}
\end{figure}

This deviation of the self-consistent single-particle potential
\begin{equation}
U_{BHF} (k) = \mbox{Real}\left[\Sigma_{BHF} (\vec k,\omega = \varepsilon_k)
\right] \label{eq:ubhf}
\end{equation}
from its quadratic parametrisation, which is implied by (\ref{eq:param1}), is
explicitly displayed in Fig.~\ref{fig2}, where results are given at two
different densities using various models for the NN interaction. These models
include the versions Bonn A and Bonn C of the traditional One-Boson-Exchange
model defined in \cite{rupr}, the interaction models with high accuracy fits to
the NN data CD Bonn\cite{cdb} and Argonne V18\cite{argv18}, but also the recent
NN potential Idaho A\cite{idaho}, which is based on chiral perturbation theory.

All these single-particle potentials show a significant deviation from a
parabolic shape in particular at momenta slightly above the Fermi momentum. It
is obvious that such a deviation tends to provide more attractive matrix
elements of $G$ in evaluating the self-energy for hole states according
(\ref{eq:selfbhf}), which leads to more binding energy. 

\begin{figure}
\begin{center}
\epsfig{figure=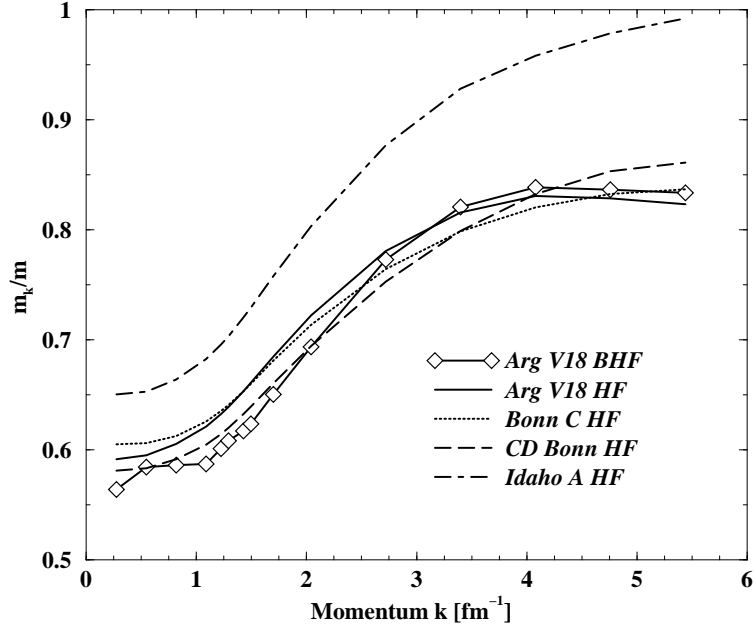,width=10cm}
\end{center}
\caption{The effective k-mass (see eq.(\protect\ref{eq:kmass})) evaluated for
the BHF and Hartree-Fock (HF) self-energy for various interactions. Symmetric
nuclear matter at the empirical saturation density ($k_F$ = 1.36 fm$^{-1}$) has
been considered.\label{fig3}}
\end{figure}

In the following we want to explore the sources of the momentum dependence of
$U_{BHF}$ more in detail. For that purpose we have disentangled the
dependence of the real part of the self-energy $\Sigma_{BHF} (\vec k,\omega)$
on energy and momentum. Following the nomenclature of Mahaux and
Sartor\cite{mahaux} we characterise the dependence on the momentum $k$ by an
effective ``k-mass'' $m_k$, which is defined by
\begin{equation}
\frac{m_k(k)}{m} = \left[1 + \frac{m}{k}\frac{\partial\Sigma_{BHF}(k,\omega)}
{\partial k}\right]^{-1}\,.\label{eq:kmass}
\end{equation}
Any deviation of $m_k$ from the bare mass $m$ indicates a momentum-dependence of
the self-energy, which means a non-locality in coordinate space. Results for the
BHF self-energy calculated for the Argonne V18 calculated in nuclear matter at
a Fermi momentum $k_F$ = 1.36 fm$^{-1}$ are displayed in Fig.~\ref{fig3}. For
comparison we also include in this figure the effective k-mass evaluated in the
Hartree-Fock approximation, i.e.~replacing the matrix elements of $G$ in
(\ref{eq:selfbhf}) by the corresponding ones of the bare interaction $V$. One
can see that the k-mass evaluated in the BHF approximation is very close to
the one determined from the Hartree-Fock potential. This means that the
non-locality of the single-particle potential originates essentially from the
Hartree-Fock contribution. 

The same is true for other interactions considered in this investigation.
Therefore using other NN potentials we only display the effective k-mass 
derived from the Hartree-Fock approach in Fig.~\ref{fig3}. It is remarkable
that the effective k-mass derived for the traditional Bonn interaction as well 
as for the CD Bonn potential shows results very close to the ones derived from
the V18 interaction model. This is true although the absolute values for the
Hartree-Fock single-particle potentials depend quite strongly on the NN
interaction, as one can see from the single-particle energies listed in
the first column of table~\ref{tab1}. This means that different interaction
models, which all fit NN scattering data, yield quite different values for the
Hartree-Fock single-particle potential in nuclear matter. The momentum
dependence of these single-particle potentials, which represents the non-locality, is
very similar. 

In Fig.~\ref{fig3} it is only the Idaho A interaction model\cite{idaho}, 
as well as the version B of this model which is not contained in this figure,
which shows significant deviations in particular at high momenta. These
differences are due to the strong cut-off, which are employed in these
interaction models. These cut-offs, which are necessary the chiral expansion at
high momenta, are responsible for the fact that the Hartree-Fock potential is 
essentially identical to zero for momenta larger than 5 fm$^{-1}$.

\begin{figure}
\begin{center}
\epsfig{figure=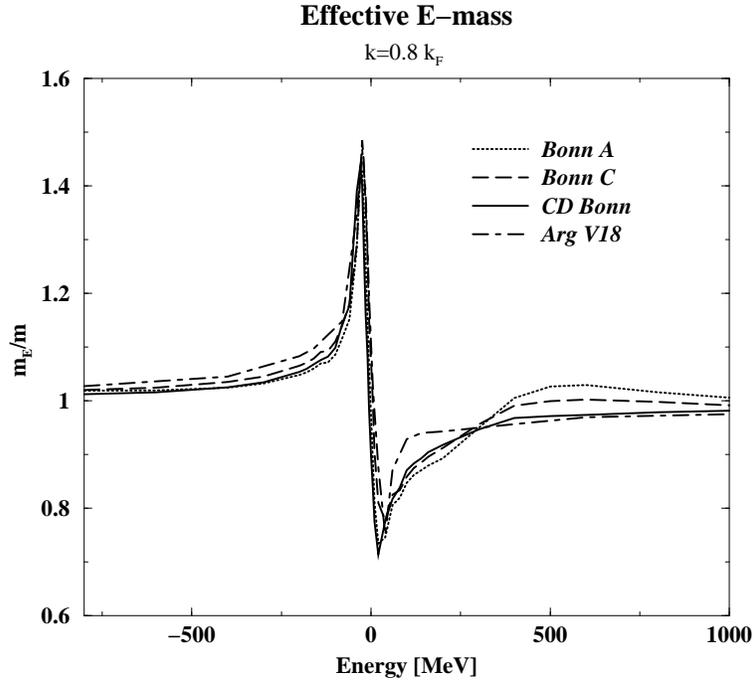,width=10cm}
\end{center}
\caption{The effective E-mass (see eq.(\protect\ref{eq:emass})) evaluated for
the BHF self-energy for various interactions. Symmetric
nuclear matter at the empirical saturation density ($k_F$ = 1.36 fm$^{-1}$) has
been considered, using as a typical example a momentum of $k=0.8\ k_F$.
\label{fig4}}
\end{figure}

All k-masses exhibit a rather smooth dependence on the momentum. Therefore the
momentum dependence of the self-energy is not responsible for the special
behaviour of the BHF self-energies at the $k=k_F$ displayed in
Fig.~\ref{fig2}. The energy dependence of the self-energy, so to say the
non-locality in time, is characterised by
the so-called ``E-mass'', which is defined as\cite{mahaux}
\begin{equation}
\frac{m_E(\omega)}{m} = \left[1 - \frac{\partial\Sigma_{BHF}(k,\omega)}
{\partial \omega}\right]\,,\label{eq:emass}
\end{equation}
so that the total effective mass is given by
$$
\frac{m^*(k)}{m} = \frac{m_k(k)}{m}\frac{m_E(\omega=\varepsilon_k)}{m}\,.
$$
In the Hartree-Fock approximation the E-mass is identical to $m$ for all NN
interaction models considered in this investigation. Therefore the deviations of
$m_E$ from $m$, which are displayed in Fig.~\ref{fig4} for various NN 
interactions originate from the ladder contributions to the G-matrix. Again we
find a behaviour which is very similar for all interactions considered.  
The effective E-mass reaches values up to 1.4 times $m$ at energies $\omega$
slightly above the Fermi energy, which implies that the total effective mass
$m^*$ approaches the value $m$ for momenta around the Fermi momentum, indicating
that the BHF self-energy increases only weakly with $k$, just the behaviour we
already observed in Fig.~\ref{fig2}. 

\begin{figure}
\begin{center}
\epsfig{figure=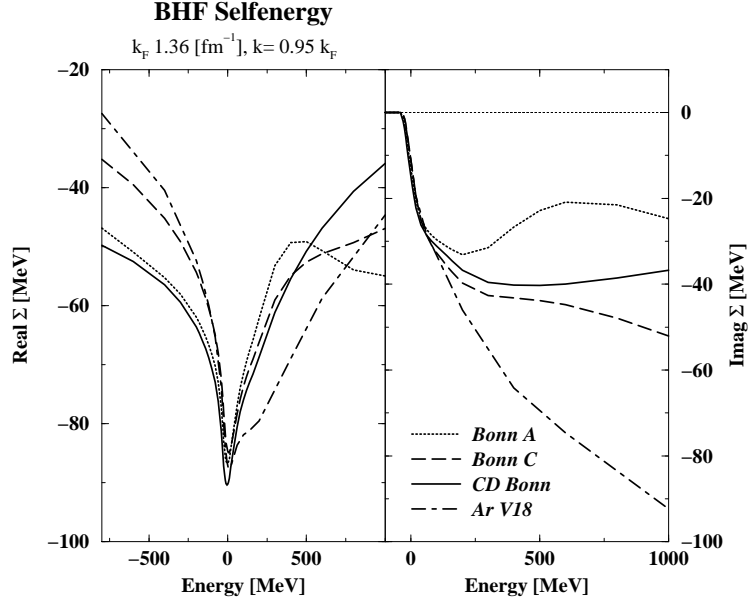,width=10cm}
\end{center}
\caption{The BHF self-energy (see \protect{\ref{eq:selfbhf}}) for symmetric 
nuclear matter with Fermi momentum $k_F$ = 1.36 fm$^{-1}$ calculated for
nucleons with momentum $k=$ 0.95 $k_F$ as a function of the energy $\omega$. 
The real and imaginary part of the
self-energy are displayed in right and left part of the figure, respectively.
Various realistic NN interactions have been considered, which can be
distinguished by the line-types as listed in the figure.
\label{fig5}}
\end{figure}

This energy dependence of the real part of the self-energy is also visualised 
in the left part of Fig.~\ref{fig5} displaying a pronounced minimum at energies
around the Fermi energy. The right hand part of this figure shows the
corresponding imaginary parts of the BHF self-energy. These imaginary parts are
identical zero for energies $\omega$ less than $\varepsilon_k - \varepsilon_F$,
as can be seen from eqs.(\ref{eq:bhf1}) and (\ref{eq:bhf2}), and yield
non-negligible values up to very high energies.

Real and imaginary part of the self-energy are related to each other by a
dispersion relation of the form\cite{mahaux,wimo}
\begin{equation}
\mbox{Real}\,\Sigma_{BHF}(k,\omega) = U_{HF}(k) + \frac{1}{\pi}
\int_{-\infty}^\infty \frac{\mbox{Imag}\,
\Sigma_{BHF}(k,\omega')}{\omega'-\omega}d\omega'\,.\label{eq:disper1}
\end{equation}
Considering the results for the imaginary part of the self-energy, which are
displayed in the right part of Fig.~\ref{fig5}, it is clear from this
dispersion relation that the real part of the self-energy is identical to the
corresponding HF single-particle potential $U_{HF}(k)$ in the limit $\omega \to
-\infty$. This real part gets more attractive with increasing $\omega$ until
one reaches  values of $\omega$ at which the imaginary part is different from
zero. The self-energy turns less attractive at higher energies, which leads to
a pronounced minimum at energies $\omega$ slightly above the Fermi energy. This
energy dependence is of course also reflected in the effective $E$-mass 
discussed above. 

\begin{figure}
\begin{center}
\epsfig{figure=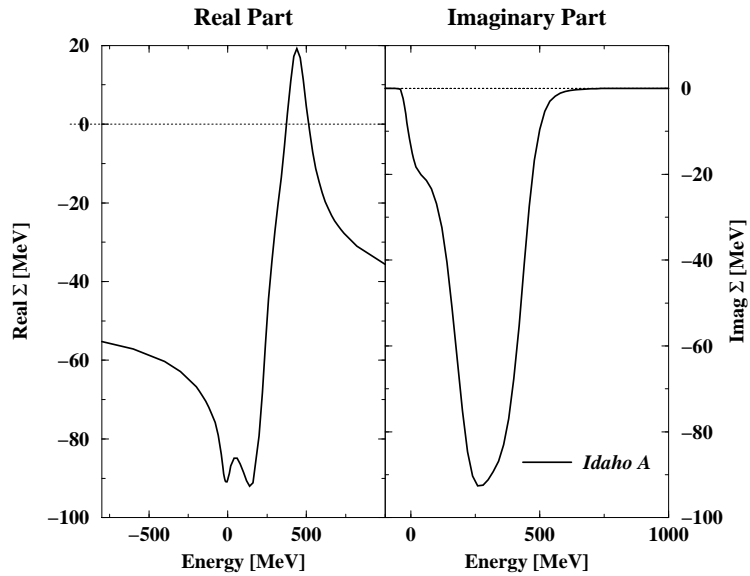,width=10cm}
\end{center}
\caption{The BHF self-energy (see \protect{\ref{eq:selfbhf}}) evaluated for the
Idaho A interaction model. Further details as in Fig.~\protect{\ref{fig5}}
\label{fig6}}
\end{figure}

The results obtained for the recent Idaho interactions $A$ and $B$ are quite
different as can be seen from Fig.~\ref{fig6}, which displays corresponding
results for the BHF self-energy using Idaho A interaction. These differences
can be traced to the strong cut-off of these Idaho interactions which have been
mentioned already above. Due to these formfactors the matrixelements 
$<\vec k\vec q | V |\vec p_1 \vec p_2 >$ in the Bethe-Goldstone
eq.(\ref{eq:betheg}) vanish for two-body states $|\vec p_1 \vec p_2 >$ with high
momenta $\vec p_i$. Therefore the imaginary part of the self-energy tends to
zero for energies $\omega$ larger than 500 MeV as shown in the left part of
Fig.~\ref{fig6}. The dispersion relation of (\ref{eq:disper1}) relates this
shape of the imaginary part to the energy-dependence of the real part of
$\Sigma_{BHF}$ leading even to positive values for $\omega \approx$ 400 MeV.
It is worth noting that the large absolute value of imaginary part of
$\Sigma_{BHF}$ obtained for the Idaho interaction at energies below 250 MeV
yields a real part of the self-energy which is similar to the results obtained
for other interactions in the energy interval for $\omega$ of [-100 MeV, 0],
which is relevant for  the BHF single-particle energies. Therefore one can
expect that these Idaho interactions lead to reasonable predictions for nuclear
structure at low energies and momenta. The results for processes involving 
nucleons with high momenta or energies will be dominated by the strong cut-off,
which is required in these models to control the expansion of the chiral
perturbation theory. 

The different NN interactions yield quite different results for the HF
single-particle potential but very similar ones for the BHF self-energy at
energies $\omega$ which are close to the Fermi energy (see Fig.~\ref{fig5} and
Table \ref{tab1}). We can distinguish ``stiff'' potentials like the Argonne
potential V18 or the Bonn C potential and ``softer'' ones like Bonn A and CD
Bonn or the even softer ones Idaho A and B. The stiff potentials yield rather
repulsive results for mean field of the Hartree-Fock approximation (see first
column of table~\ref{tab1}). This repulsion originating from the bare NN
interaction $V$ is compensated by the attractive particle-particle ladder
contributions to the $G$-matrix. The value of this attractive contribution
depends on the interaction used. It is smaller for softer potentials. If
the two-particle propagator in the Bethe-Goldstone eq.~(\ref{eq:betheg}) was
replaced by the free one,
i.e.~no Pauli operator $Q$ and the energies in the denominator replaced by the
kinetic energies, the $G$ would become identical to the T-matrix of free NN
scattering. Since all potentials fit the same NN phase shifts a replacement of
$G$ in the BHF equations by T should yield identical results. Pauli operator and
the single-particle energies in the medium lead to less attractive matrix
elements of $G$ as compared to $T$, which is often called Pauli- and dispersion
quenching, respectively. This quenching mechanism is more efficient for the
stiff potentials, since the attractive ladder contributions, which are quenched,
are larger. Therefore stiff potentials lead to smaller binding energies in BHF
calculations of nuclear matter as softer ones. 

This can be seen from columns 1 to 3 in Table~\ref{tab2}, which lists results
for the binding energy obtained in different BHF calculations of nuclear matter
at three different densities, employing various realistic NN interactions. The
dispersion quenching mechanism tends to be larger if one employs the
conventional choice (\ref{eq:conv}) with a gap in the single-particle spectrum
for the Bethe-Goldstone equation as compared to the continuous choice. Therefore
one obtains larger binding energies using the continuous choice. This is can be
seen in the comparison between columns 1 and 2 of Table~\ref{tab2}. In both
kinds of calculations a parametrisation of the single-particle spectrum
according to (\ref{eq:param1}) has been utilised. An even larger binding energy
is obtained if the exact Pauli operator and the precise shape of the
single-particle energies are employed (see third column in Table~\ref{tab2}).
As discussed above this is partly due the use of the exact Pauli operator, but
it is also due to the shape of the single-particle energy at $k=k_F$ (see
Fig.~\ref{fig2}), which is characterised by a large effective E-mass at these
momenta (see Fig.~\ref{fig4}).

\section{Towards a self-consistent single-particle Green function}

One of the arguments in favour of the continuous choice to be used in the
single-particle spectrum of the Bethe-Goldstone equation has been that the
corresponding definition of the single-particle self-energy would be in line
with the definition of the self-energy using the method of self-consistent Green
functions\cite{mupo,wimo}. Assuming that the self-energy $\Sigma(k,\omega)$ for
a nucleon with momentum $k$ and energy $\omega$ in infinite nuclear matter is
given, the Dyson equation leads to a single-particle Green function of the form
\begin{equation}
g(k,\omega)  = \frac{1}{\omega - \frac{k^2}{2m} - \Sigma(k,\omega)}\,.
\label{eq:green1}
\end{equation}
If one compares this solution with the general Lehmann representation
\begin{equation}
g(k,\omega) =  {\lim_{\eta\to 0}} \left( \int_{-\infty}^{\epsilon_{\rm F}}
d\omega' \frac{S_{\rm h}(k,\omega')}{\omega -\omega' - i\eta} +
\int_{\epsilon_{\rm F}}^{\infty}
d\omega' \frac{S_{\rm p}(k,\omega')}{\omega -\omega' + i\eta}
\right)\, ,
\label{eq:lehman}
\end{equation}
one can easily identify the spectral functions $S_h(k,\omega)$ and
$S_p(k,\omega)$ for hole and particle strength, respectively, to be given as
\begin{equation}
S_{h(p)}(k,\omega)= \pm \frac {1}{\pi} \frac {\mbox{Imag} \Sigma(k,\omega)}
{(\omega -k^2/2m - \mbox{Real} \Sigma(k,\omega))^2 + (\mbox{Imag} \Sigma (k,
\omega))^2} ~,~ \mbox{for} \ \omega < \varepsilon_F ~ (\omega >
\varepsilon_F)\,.
\label{eq:spectf}
\end{equation}
The hole spectral function represents the probability that a particle with
momentum $k$ and energy $\omega$ can be removed from the ground state of the
system, leaving the residual nucleons in an eigenstate of the hamiltonian. The
particle spectral function system contains the corresponding probability for
adding a particle. In the mean field or Hartree-Fock approximation these
spectral functions are reduced to $\delta$-functions of the form
\begin{eqnarray}
S_p(k,\omega) & = & \Theta(k - k_F) \delta(\omega - \varepsilon_k)\nonumber\\
S_h(k,\omega) & = & \Theta(k_F - k) \delta(\omega -
\varepsilon_k)\,,\label{eq:spec0}
\end{eqnarray}
where the single-particle energy $\varepsilon_k$ contains the sum of kinetic
energy plus the real part of the self-energy $\Sigma(k,\omega)$ calculated
on-shell ($\omega = \varepsilon_k$). 

The two-particle propagator can then be written in the form
\begin{eqnarray}
g_{II}(k_1,k_2;\Omega)  & = & \int_{\varepsilon_F}^\infty d\omega_1
\int_{\varepsilon_F}^\infty d\omega_2 \frac{S_p(k_1,\omega_1)S_p(k_2,
\omega_2)}{\Omega-\omega_1-\omega_2+i\eta} \nonumber\\ && \qquad
- \int_{-\infty}^{\varepsilon_F}
d\omega_1\int_{-\infty}^{\varepsilon_F} d\omega_2 \frac{S_h(k_1,\omega_1)
S_h(k_2,\omega_2)}{\Omega-\omega_1-\omega_2-i\eta}\label{eq:green2}\,.
\end{eqnarray}
If one approximates the spectral functions by the mean field approximation one
obtains the Galitskii-Feynman propagator, which includes particle-particle
propagation as well as hole-hole propagation. If the hole-hole part of the
propagator is ignored one obtains an equation for the ladder diagrams for the
reducible two-particle Green function which corresponds to the Bethe-Goldstone
equation employing the continuous choice. Using the complete Galitskii-Feynman 
propagator for nuclear matter at temperature $T=0$ with a realistic NN 
interaction leads to the so-called pairing- or
deuteron-instability\cite{vonder}.   

As a side remark we would like to mention that this instability has also been
observed in studying correlation energies in finite nuclei\cite{mavro}. The
attempt to solve the particle-particle hole-hole RPA equations for finite nuclei
in a large model space using realistic interactions and BHF single-particle
energies leads to complex eigenvalues of the RPA equation. This
corresponds to the pairing instability in the Galitskii-Feynman equation for
infinite matter mentioned above. In finite nuclei this instability could be
removed by a self-consistent single-particle propagator, which goes beyond the
BHF approximation and accounts for the distribution of single-particle
strength\cite{heinz}. 

Guided by this experience in finite nuclei one can expect that the pairing
instability could be removed if the single-particle Green function are evaluated
in a way which account for the spectral distributions in a self-consistent way.
In fact, some attempts in this direction are under investigation and first
results are rather promising\cite{gearh,roth}. In this investigation we would
like to follow a different route and calculate the contribution of the
hole-hole terms to the self-energy in a kind of perturbative way\cite{gcl}
\begin{equation} 
\Delta \Sigma_{2h1p} (k,\omega) = \int_{k_F}^\infty d^3p \int_0^{k_F}
d^3h_1\,d^3h_2\, \frac{<k,p\vert G\vert h_1,h_2>^2}{\omega +
\tilde\varepsilon_p - \tilde\varepsilon_{h_1} -
\tilde\varepsilon_{h_2}-i\eta}\,.\label{eq:2h1p}
\end{equation}
In a first approximation, which we will denote as Extended Brueckner - Hartree -
Fock 1 (EBHF1), we assume a single-particle spectrum $\tilde\varepsilon_k$ which
Illegal variable name.
is identical to the self-consistent BHF spectrum, discussed above
(\ref{eq:bhf1}), but shifted
by a constant $C_1$, which ensures the self-consistency for $k=k_F$
\begin{eqnarray}
\tilde\varepsilon_{k_F} & = & \varepsilon_{k_F}^{BHF} + C_1 \nonumber\\
& = & \frac{k_F^2}{2m} +  \Sigma_{BHF} (k_F,\omega = \tilde\varepsilon_{k_F}) + 
\Delta \Sigma_{2h1p} (k_F,\omega= \tilde\varepsilon_{k_F})\,.\label{eq:ebhf1}   
\end{eqnarray}
This shifted single-particle spectrum is also used in the Bethe-Goldstone
equation.

\begin{figure}
\begin{center}
\epsfig{figure=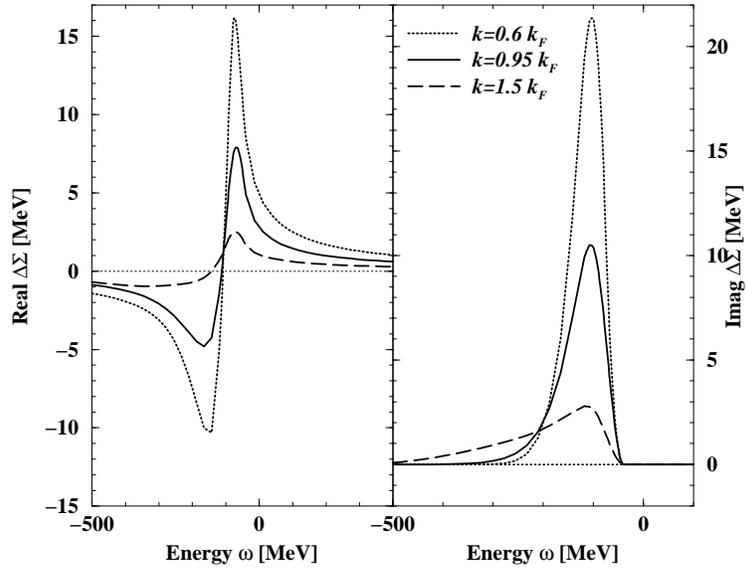,width=10cm}
\end{center}
\caption{The 2h1p contribution to the self-energy (see \protect{\ref{eq:2h1p}}) 
evaluated for the CD-Bonn interaction assuming $k_F$ = 1.36 fm$^{-1}$ calculated 
for various momenta $k$. Real (left part) and imaginary part(right part of the
figure) are displayed as function of $\omega$. 
\label{fig7}}
\end{figure}

Results for the two-hole one-particle contribution to the self-energy, $\Delta 
\Sigma_{2h1p}$, are displayed in Fig.~\ref{fig7}, considering various momenta
$k$. The imaginary part of $\Delta \Sigma_{2h1p}$ is different from zero only
for energies $\omega$ below the Fermi energy. The conservation of the
total momentum in the two-nucleon of the G-matrix in (\ref{eq:2h1p}), $\vec h_1
+ \vec h_2 = \vec k + \vec p$, leads to a minimal value of $\omega$ at which
this imaginary part is different from zero. Due to these limitations the
imaginary part integrated over all energies is much smaller for $\Delta
\Sigma_{2h1p}$ than for $\Sigma_{BHF}$, displayed in Fig.~\ref{fig5}. The real
part of $\Delta \Sigma_{2h1p}$ is related to the imaginary part by a dispersion
relation similar to the one of (\ref{eq:disper1}), connecting the imaginary part
of $\Sigma_{BHF}$ with the particle-particle ladder contributions to the real
part of $\Sigma_{BHF}$. Since the imaginary part of $\Sigma_{2h1p}$ is
significantly smaller than the one of $\Sigma_{BHF}$, the same is true also for
the corresponding real part. 

This is reflected in the single-particle energies listed in Table~\ref{tab1}.
While the contribution of the particle-particle ladders to the Fermi energy is
as large as -90 MeV for the Bonn C or Argonne V18 interaction, the correction
originating from the 2h1p term is only around 5 MeV. This is a justification of
the perturbative treatment of the hole-hole ladder contribution. The
non-symmetric treatment of particle-particle and hole-hole excitations, is the
typical feature of the hole-line expansion. The hole-line expansions seems to be
justified for nuclear systems at densities around the saturation density if one
is using realistic NN interaction. The origin for this difference in the
importance of particle-particle and hole-hole excitations can be read from a
comparison of the imaginary parts of the self-energy contributions displayed in
Fig.~\ref{fig5} and \ref{fig7}. The phase space of particle-particle
excitations, which can be excited from the ground-state by a realistic NN
interaction, which is represented by an imaginary part of $\Sigma_{BHF}$
different from zero, covers a much larger energy interval than the
corresponding space of hole-hole excitations (see imaginary part of $\Delta
\Sigma_{2h1p}$). The relative importance of the hole-hole contributions shall be
larger at very high densities as the phase space of hole-hole configurations
increases. The relative importance of hole-hole contributions is also larger for
interactions like the Idaho A and Idaho B model, for which the particle-particle
configurations which can be reached by the interaction are strongly limited by
the form factors used (see Fig.~\ref{fig6} and Table~\ref{tab1}).

\begin{figure}
\begin{center}
\epsfig{figure=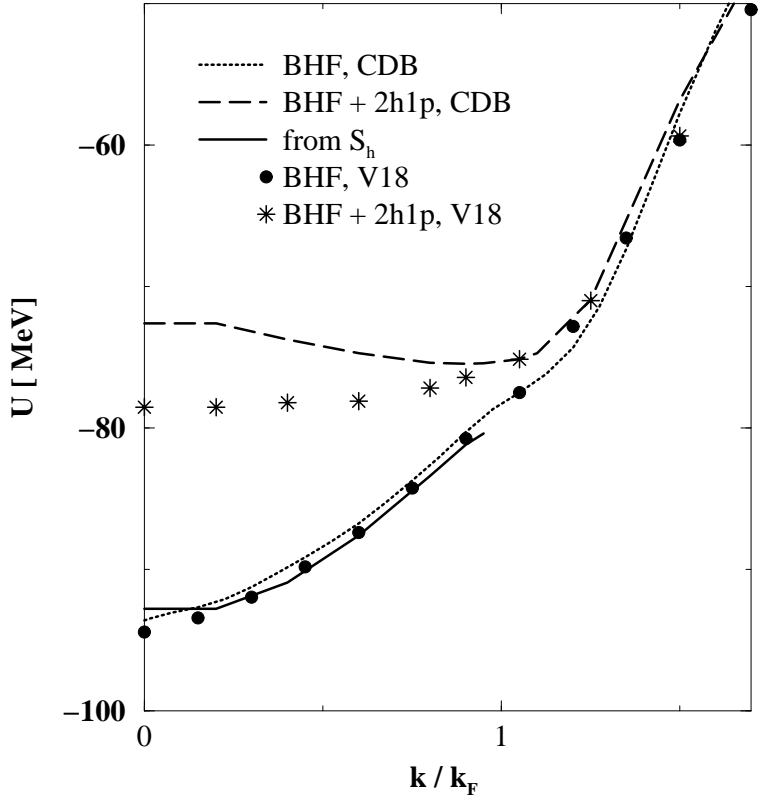,width=10cm}
\end{center}
\caption{The single-particle potential, i.e. the single-particle energy minus
the kinetic energy, as a function of the momentum assuming the BHF
approximation, the quasiparticle energy originating from the BHF + 2h1p (EBHF1)
approximation and the mean value derived from the hole spectral function
according to (\protect{\ref{eq:meaneh}}). 
The data have been obtained for nuclear matter with a Fermi
momentum $k_F$ = 1.36 fm$^{-1}$, assuming the CD Bonn potential and the Argonne
V18. The results for the V18 interaction have been shifted by a constant such
that the BHF results for $k=k_F$ agree with those derived from the CD Bonn
potential. \label{fig8}}
\end{figure}

The corrections of the single-particle energies due to the $\Delta\Sigma_{2h1p}$
term are larger for momenta below the Fermi momentum and tend to zero for
momenta above $k_F$. This can be seen already from Fig.~\ref{fig7} and is shown 
explicitely in Fig.~\ref{fig8}, where the single-particle potential, i.e.~the
quasiparticle energy
\begin{equation}
\varepsilon_{qp}(k) = \frac{k^2}{2m} + \Sigma_{BHF} (k_F,\omega =
\varepsilon_{qp}(k)) + 
\Delta \Sigma_{2h1p} (k_F,\omega= \varepsilon_{qp}(k))\,,\label{eq:ebhf1a}
\end{equation}
minus the kinetic energy is compared to the corresponding value obtained in the
BHF approximation. One finds that the single-particle potential derived from the
quasiparticle energy tends to a constant for momenta below $k_F$. This means
that the effective mass which describes the momentum dependence of the
quasiparticle energy $\varepsilon_{qp}$ is essentially equal to the bare mass
for all momenta below $k_F$. At first sight one may be tempted to consider this
quasiparticle spectrum also in the energy denominators of the Bethe-Goldstone
equation and the 2h1p correction term of eq. (\ref{eq:2h1p}). We will see below,
however, that a different choice is more appropriate.

It is worth noting that the momentum dependence of the single-particle spectra,
in the BHF as well as in the EBHF1 approximation, is very similar for all
interactions considered. This means that the values obtained for $k$ smaller
than the Fermi momentum using the various 
interactions deviate by less than 1 MeV if they are normalised relative to the
corresponding Fermi energy. Only the  $\Sigma_{2h1p}$ correction term is about
25 percent weaker for the Argonne V18 than for the various Bonn or Idaho
interaction models (see also Fig.~\ref{fig8}).

\begin{figure}
\begin{center}
\epsfig{figure=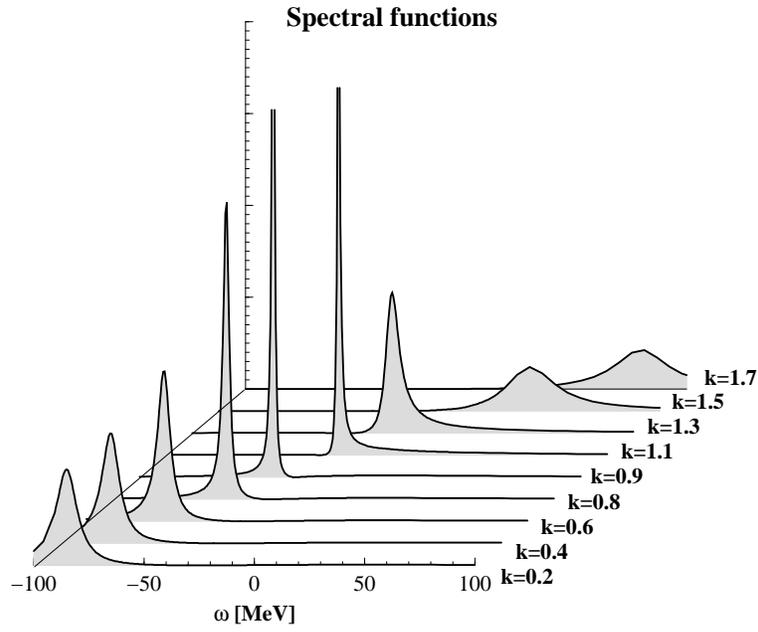,width=10cm}
\end{center}
\caption{The spectral function for particle and hole strength
$S_h(k,\omega)+S_p(k,\omega)$ as a function of energy assuming various momenta.
The data
have been obtained for nuclear matter with a Fermi momentum $k_F$ = 1.36
fm$^{-1}$, assuming the CD Bonn potential.  
\label{fig9}}
\end{figure}

From the complex self-energy $\Sigma(k,\omega)$ in the EBHF1 approach,
which is the sum of the BHF term and the 2h1p correction, one can evaluate the
single-particle Green function or directly the spectral functions according to
(\ref{eq:spectf}).  Results for the spectral function derived from the CD Bonn
interaction for nuclear matter at the empirical saturation density are displayed
in Fig.~\ref{fig9}. For each momentum $k$ the spectral functions exhibit a 
maximum at the quasiparticle energy $\varepsilon_{qp}(k)$. The width of this
maximum is very small for $k$ close to the Fermi momentum and gets significantly
larger for very small momenta and momenta considerably larger than $k_F$. 

\begin{figure}
\begin{center}
\epsfig{figure=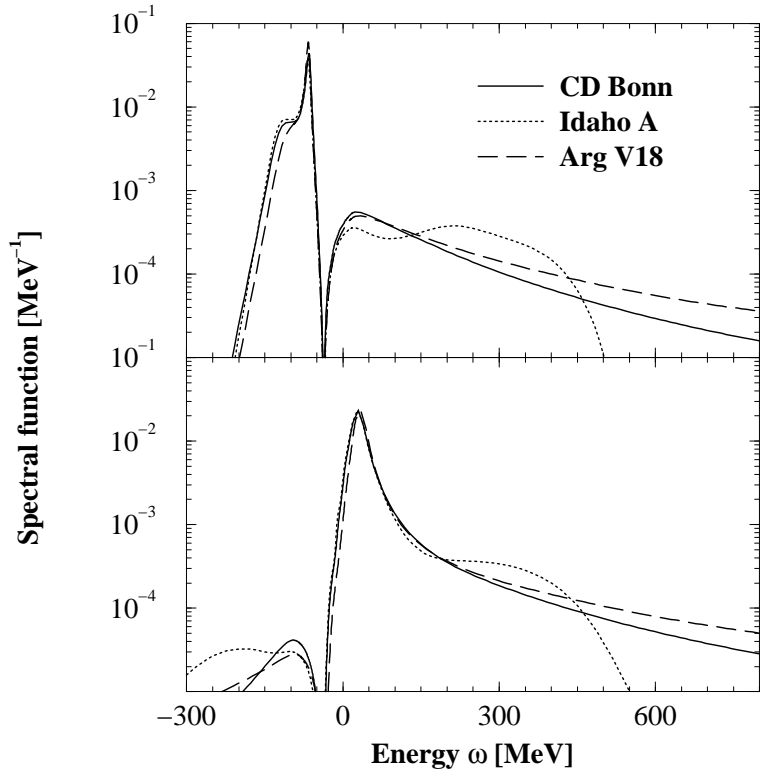,width=10cm}
\end{center}
\caption{The spectral function for particle and hole strength
$S_h(k,\omega)+S_p(k,\omega)$ as a function of energy assuming $k=0.4\ k_F$
(upper part of the figure) and $k=1.5\ k_F$ (lower half of the figure).
The data
have been obtained for nuclear matter with a Fermi momentum $k_F$ = 1.36
fm$^{-1}$, assuming three different NN interactions.  
\label{fig10}}
\end{figure}

More details can be seen in the logarithmic plots of the spectral function in
Fig.~\ref{fig10}. This representation also exhibits some characteristic
differences depending on the interaction used. While the spectral functions
derived from CD Bonn and Argonne V18 interaction exhibit a high-energy tail
which extends to excitation energies $\omega$ above 1 GeV, the spectral function
determined for the Idaho interaction drop very sharply at energies around 400 to
500 MeV. This is again a consequence of the strong cut-offs which are used in
these interactions to control the terms in the chiral perturbation expansion.

From this figure one also observes of course that the spectral distributions are
not symmetric around the quasiparticle pole. To demonstrate this on a
quantitative level we calculate e.g.~the mean value for the energy of the hole 
distribution function
\begin{equation}
\bar{\varepsilon_h}(k) = \frac{\int_{-\infty}^{\varepsilon_F} d\omega\,\omega\,
S_h(k,\omega)}{n(k)}\,,\label{eq:meaneh}
\end{equation}
where $n(k)$ denotes the occupation probability for the  state with momentum
$k$, which is calculated as
\begin{equation}
n(k) =  \int_{-\infty}^{\varepsilon_F} d\omega\,S_h(k,\omega)\,.\label{eq:n(k)}
\end{equation}
These mean values are significantly below the quasiparticle energies (see solid
line in Fig.~\ref{fig8}). In fact, for all interactions and densities under
consideration it turned out that these results for $\bar{\varepsilon_h}(k)$
are close to the BHF single-particle energies for momenta $k\leq k_F$.

\begin{figure}
\begin{center}
\epsfig{figure=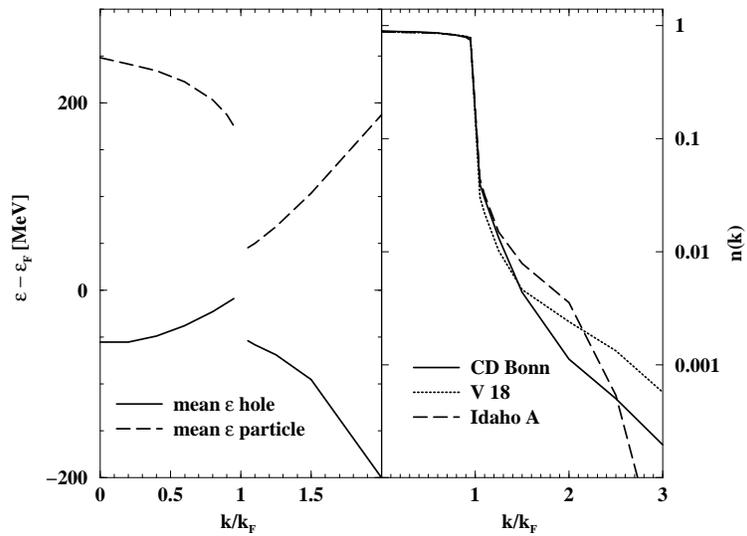,width=10cm}
\end{center}
\caption{The left part of the figure exhibits the mean value for the
single-particle energy weighted by the spectral function for particle and hole 
strength, respectively, relative to the Fermi energy. Results are given for the
example of the Bonn C potential. The part on the right-hand side shows the
occupation probability $n(k)$ for various interaction. All results in this
figure refer to nuclear matter with a Fermi momentum $k_F$ = 1.36
fm$^{-1}$.  
\label{fig11}}
\end{figure}

However, the mean value $\bar{\varepsilon_h}(k)$ is defined also for momenta
larger than $k_F$. For those momenta we do not get a dominant contribution from
the quasiparticle pole, but determine an average over a broad distribution of
2h1p configurations. Therefore the mean energies $\bar{\varepsilon_h}(k)$ are
much more attractive for $k > k_F$ than for $k < k_F$ as it is shown in
Fig.~\ref{fig11}. As a consequence the total energy per nucleon calculated as
\begin{equation}
\frac{E}{A} = \frac{\int d^3k\,\int_{-\infty}^ {\varepsilon_F} d\omega\,
S_h(k,\omega)\frac{1}{2}\left(\frac{k^2}{2m} + \omega\right)}{\int d^3k\,n(k)}
\,,\label{eq:ebhf2}
\end{equation} 
is significantly more attractive than the corresponding BHF result (see
Table~\ref{tab2}, column denoted EBHF1 as compared to BHF).

Fig.~\ref{fig11} displays in its right part also the momentum distribution
$n(k)$ derived from the CD Bonn interaction (the momentum distribution for the
other two Bonn potential are very similar), the Argonne V18 and the Idaho A
interaction. At high momenta the result is larger for the Argonne V18 as
compared to the Bonn interaction model. The Idaho interaction predicts a
momentum distribution that decreases very rapidly at momenta larger than 2
$k_F$. This is again a consequence of the strong cut off in this interaction and
should not be considered as a realistic prediction.
 
The left part of Fig.~\ref{fig11} also shows results for the mean energy of the
particle strength distribution
\begin{equation}
\bar{\varepsilon_p}(k) = 
\frac{\int_{\varepsilon_F}^{\infty} d\omega\,\omega\,
S_p(k,\omega)}
{\int_{\varepsilon_F}^{\infty} d\omega\,
S_p(k,\omega)}\,\label{eq:meanep}
\end{equation}
In this integration one has to include strength up to energies $\omega$ above 1
GeV, to ensure that the sum-rule
$$
n(k) + \int_{\varepsilon_F}^{\infty} d\omega\,S_p(k,\omega) = 1
$$ 
is fulfilled. 

For $k$ smaller than the Fermi momentum more than 80 percent of the
single-particle strength is at energies below the Fermi energy. This means that
the single-particle energy for these momenta might be represented by 
$\bar{\varepsilon_h}(k)$. On the other hand, for $k$ larger than $k_F$, the
dominant part of the single-particle strength is represented by the mean energy
$ \bar{\varepsilon_p}(k)$. Therefore one might use these energies for the
single-particle spectrum to be used in the propagator of the Bethe-Goldstone
equation and the energy denominator of $\Delta \Sigma_{2h1p}$ in
(\ref{eq:2h1p}). As one can see from Fig.~\ref{fig11} this would lead to a
single-particle spectrum of 40 to 50 MeV at the Fermi momentum, which is
more than one half of the gap in the conventional choice for BHF. 

Instead of this choice we suggest a slightly different one. Since we would like
to use these mean energy values to define an approximation to the
single-particle Green function (\ref{eq:lehman}) we define a mean value
$\hat\epsilon_h(k)$ for hole states by the equation
\begin{equation}
\frac{1}{\hat\epsilon_h(k)-\left(\varepsilon_F + \frac{\Delta}{2}\right)} =
\frac{1}{n(k)}\int_{-\infty}^{\varepsilon_F} d\omega \frac{S_h(k,\omega)}{\omega 
-\left(\varepsilon_F + \frac{\Delta}{2}\right)}\label{eq:EBHF22}
\end{equation}
and a corresponding one for the particle states. With this definition of a mean
value one reduces in particular the contributions to the mean value for the
particle states which originates from the spectral strength $S_p(k,\omega)$ at
very high energies $\omega$ in an appropriate way. 

\begin{figure}
\begin{center}
\epsfig{figure=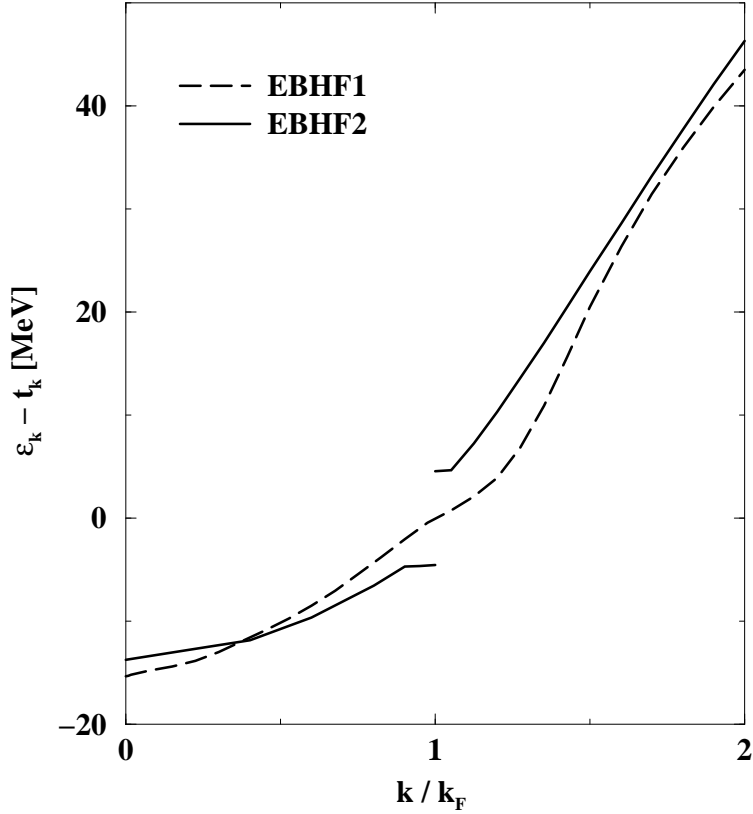,width=10cm}
\end{center}
\caption{The mean value $\hat\epsilon_h(k)$ defined in
eq.(\protect\ref{eq:EBHF22}) and the corresponding $\hat\epsilon_p(k)$ (solid
lines) are compared to the single-particle spectrum obtained within the BHF
approximation (dashed line). Note that the kinetic energies have been
subtracted and the curves have been shifted to obtain the value 0 for $k=k_F$. 
The results in this figure refer to nuclear matter with a Fermi momentum $k_F$
= 1.36 fm$^{-1}$ and have been derived from the CD Bonn interaction.  
\label{fig12}}
\end{figure}

Examples for these mean values are plotted in Fig.~\ref{fig12}. With this
definition of the single-particle spectrum one obtains a gap $\Delta$ at the
Fermi energy $\varepsilon_F$ of about 10 MeV for nuclear matter at saturation
density. It is worth noting that this gap is of similar size as the pairing gap
derived from BCS calculations for T=0 pairing in the $^3S_1$ - $^3D_1$
channel\cite{lomb}. This implies that using this single-particle spectrum in a
Galitzkii-Feynman equation should avoid the occurrence of pairing instabilities.
Also one may mention that such a gap is in between the
behaviour of the conventional and the continuous choice for the BHF spectrum
discussed above.

This single-particle spectrum has then be used in the Bethe-Goldstone equation
and the evaluation of the correction term $\Delta \Sigma_{2h1p}$ of
(\ref{eq:2h1p}). We consider this choice to be an optimised representation of
the single-particle propagator in terms of one pole (at energy
$\hat\varepsilon(k)$) for each momentum $k$. Further improvements would require
the representation of the single-particle Green function in terms of two or more
poles\cite{dimitr,knehr} or use the complete spectral distribution. 

The calculation using the $\hat\varepsilon(k)$ choice is denoted by EBHF2. 
The self-energies, spectral functions and total energy are calculated in 
the same
way as discussed above for the approach EBHF1. Results for the energy per
nucleon are listed in the last column of Table \ref{tab2}. Comparing the EBHF2
results with those obtained in the EBHF1 approximation one finds that the 
gap in the single-particle spectrum yields a reduction of the calculated 
binding energy of about 1 to 3 MeV per nucleon for the densities and
interactions considered in this table. It turns out that the energies calculated
in the EBHF2 approach are again close to those obtained in the BHF approximation
using the exact propagator.

Our final EBHF2 results for the energy of nuclear matter derived from the
Argonne V18 interaction are similar to the values determined by Akmal and
Pandharipande in their variational calculation\cite{akmal} using the same
interaction model. The ``softer'' interactions like the Bonn A and the CD Bonn
interaction yield larger values for the binding energy per nucleon and larger
saturation densities. Even more binding energy is predicted from the recent
Idaho interaction models. As it has been discussed above, these Idaho models,
 seem to be useful for calculations in limited model spaces only.
Nevertheless, in order to obtain a result for the saturation point which is in
agreement with the empirical data, some repulsive effects are needed, in
particular at high densities. This repulsion can be introduced ad-hoc in terms
of a three-nucleon interaction. Such an effective three-nucleon interaction may
represent the effects of the relativistic decomposition of the
self-energy\cite{walecka} or
the effects of sub-nucleonic degrees of freedom, like e.g.~the many-body effects
arising from $\Delta$ excitations of the nucleons\cite{frick}.

\section{Conclusions}

The sensitivity of the Brueckner-Hartree-Fock (BHF) approximation for the
many-body system of symmetric nuclear matter with respect to an exact treatment
of the  propagator in the Bethe-Goldstone equation has been investigated. One
finds that the precise treatment of the Pauli operator together with a
single-particle spectrum based on the real part of the self-energy for hole-
and particle-states yields a result for the binding energy per nucleon which is
larger by a non-negligible amount as compared to results obtained in standard
approximation schemes. The non-locality and energy dependence of the BHF
self-energy is discussed in detail. 

The BHF definition of the self-energy has been extended to account for the
effects of hole-hole ladders in a perturbative way. The corresponding results
for the complex self-energy, the single-particle green function and the spectral
function are discussed in detail. This leads to a definition
of a spectrum of single-particle energies, which characterises the spectral
distribution of the single-particle Green function in an average way. The
resulting single-particle spectrum exhibits a gap at the Fermi momentum, which
is of the order of the pairing gap derived from BCS calculations for T=0
pairing in the $^3S_1$ - $^3D_1$ channel. Therefore this approximation should
avoid the so-called pairing instability which occurs in the Green function
approach using the Galitzkii-Feynman propagator.

The calculations have been performed employing various models for the NN
interaction, which all fit NN scattering data. It is observed that the strong
form factors, which had to be introduced in the recent Idaho interaction
models\cite{idaho} to control the expansion based on chiral perturbation theory, 
leads to results for the non-locality 
and spectral function, which are quite different than those obtained
for the other interactions. We conclude that these new Idaho interactions should
only be used for studies which are insensitive to components in the spectral
distribution at higher energies or momenta.

Significant differences are also observed in comparing results between stiffer
interaction models, like the Argonne V18 and Bonn C potential, or softer ones
like the CD Bonn and Bonn A interaction. These differences show up in the
Hartree-Fock contribution to the binding energy, which is rather repulsive for
the stiff interactions and less repulsive for the softer ones. These differences
can also be observed in the imaginary part of the self-energy at large energies
and in the tail of the momentum distribution at high momenta.   
    
The EBHF2 approach introduced above should be considered as a good starting
point for further improvements on a self-consistent definition of the
single-particle Green function. Such improvements include the representation of
the Green function in terms of various poles for each momentum or attempts to
account for the complete spectral distribution. The present studies demonstrate
that special attention should be paid to a proper treatment of the
single-particle strength around the Fermi energy. 

These work has been supported by the European Graduate School ``Hadrons in
Vacuum, Nuclei and Stars'' (Basel - T"ubingen). One of us, Kh.G., would like to 
thank Prof. M.M. Mustafa for useful discussions.

\begin{table}[h]
\begin{center}
\begin{tabular}{c|rrr}
&\multicolumn{1}{c}{HF} &
\multicolumn{1}{c}{BHF} & \multicolumn{1}{c}{+ 2h1p}
 \\ \hline
Bonn A & 15.66 & -37.24 & -32.48 \\
Bonn C & 60.48 & -33.53 & -29.76 \\
CD Bonn & 11.44 & -40.32 & -35.93 \\
Arg. V18 &  62.36 & -34.62 & -31.74 \\
Idaho A & -7.66 & -42.54 & -38.54 \\
Idaho B & -2.97 & -41.16 & -37.45 \\
\end{tabular}
\caption{\label{tab1} Fermi energy of nuclear matter with $k_F$ = 1.36 fm$^{-1}$
calculated in the Hartree-Fock approximation (HF), the Brueckner-Hartree-Fock
approximation (BHF) and with inclusion of 2 hole - 1 particle contributions
(+2h1p) using different interaction models. All entries are given in MeV. }
\end{center}
\end{table}

\begin{table}
\begin{center}
\begin{tabular}{cc|rrrrr}
& $k_F$ [fm$^{-1}$] &
\multicolumn{1}{c}{BHF, conv} &  \multicolumn{1}{c}{BHF, cont} &
 \multicolumn{1}{c}{BHF, exact} & \multicolumn{1}{c}{EBHF1}&
\multicolumn{1}{c}{EBHF2}
 \\ \hline
& 1.20 & -10.65 & -13.25 & -14.55 & -15.57 & -14.91 \\
Bonn A & 1.36 & -13.39 & -15.99 & -17.46 & -18.99 & -18.13 \\
& 1.60 & -16.45 & -18.81 & -20.23 & -21.93 & -21.53 \\
\hline
& 1.20 & -9.37 & -11.96 & -13.40 & -14.45 & -13.53 \\
Bonn C & 1.36 & -11.19 & -13.82 & -15.55 & -17.03 & -15.84 \\
& 1.60 & -12.11 & -14.43 & -16.40 & -18.41 & -16.69 \\
\hline
& 1.20 & -11.02 & -13.79 & -15.39 & -16.19 & -15.36 \\
CD Bonn & 1.36 & -13.88 & -16.74 & -18.83 & -20.10 & -18.94 \\
& 1.60 & -17.00 & -19.73 & -22.86 & -24.71 & -22.81 \\
\hline
& 1.20 & -9.66 & -12.03 & -13.84 & -14.65 & -13.27 \\
Arg. V18 &  1.36 & -11.32 & -13.66 & -16.13 & -18.36 & -16.62 \\
& 1.60 & -11.65 & -14.12 & -16.77 & -19.46 & -17.09 \\
\hline
& 1.20 & -11.91 & -14.64 &  -16.14 & -17.21 & -16.55 \\
Idaho A & 1.36 & -15.08 & -17.95 & -20.21 & -21.82 & -20.83 \\
& 1.60 & -18.78 & -21.88 & -25.44 & -27.62 & -26.04 \\
\hline
& 1.20 & -11.17 & -14.10 & -15.72 & -16.75 & -15.88 \\
Idaho B & 1.36 & -13.87 & -17.00 & -19.46 & -20.96 & -19.71 \\
& 1.60 & -16.55 & -19.97 & -23.94 & -25.81 & -24.07 \\
\end{tabular}
\caption{\label{tab2}  Energy per nucleon for nuclear matter considering three
different densities,  $k_F$ = 1.20, 1.36 and 1.60 fm$^{-1}$. Results are
displayed for BHF, assuming the conventional choice for the spectrum of
particle states, BHF with a continuous choice parametrised by an effective mass,
BHF using the exact Pauli operator single-particle spectrum, and for the
extended BHF schemes EBHF1  as well as EBHF2 are listed for different realistic
NN interactions. All energies are given in MeV per nucleon. }
\end{center}
\end{table}

\end{document}